\documentclass[sigconf, nonacm]{acmart}

\newcommand{\toolnamenospace}{{\tt PowerScope}}

\setcopyright{none}
\renewcommand\footnotetextcopyrightpermission[1]{}
\acmConference[]{}{}{}
\usepackage[ruled,vlined,algo2e]{algorithm2e}
\usepackage{multirow}
\usepackage{wasysym}
\usepackage{caption}
\usepackage{enumitem}
\AtBeginDocument{%
  }

\usepackage{booktabs}
\usepackage{subcaption}
\usepackage{amsmath}
\usepackage{xcolor}
\usepackage{graphicx}
\usepackage{tcolorbox}
\usepackage{adjustbox}
\usepackage{tikz}
\usetikzlibrary{arrows.meta, positioning}
\usepackage{tabularx}
\graphicspath{{figures/}}

\begin{document}

\title{\toolnamenospace: ML-based Intra-Cycle Power Estimation}


\author{Jayanth Balasubramanian}
\affiliation{%
  \institution{School of ECE, Purdue University}
  \city{West Lafayette}
  \country{USA}}
\email{jbalasub@purdue.edu}

\author{Sujay Pandit}
\affiliation{%
  \institution{School of ECE, Purdue University}
    \city{West Lafayette}
  \country{USA}}
\email{pandit8@purdue.edu}

\author{Radha Vaidya}
\affiliation{%
  \institution{Qualcomm Inc., San Diego}
  \city{San Diego}
  \country{USA}}
\email{rvaidya@qti.qualcomm.com}

\author{Anand Raghunathan}
\affiliation{%
  \institution{School of ECE, Purdue University}
    \city{West Lafayette}
  \country{USA}}
\email{raghunathan@purdue.edu}







\begin{abstract}
Power estimation at sub-clock-cycle temporal resolutions is critical for tasks such as power delivery network (PDN) design, dynamic voltage droop analysis, and pre-silicon power side-channel security evaluation. Designers commonly rely on commercial post-layout gate-level power analysis tools for these tasks, but these flows are computationally expensive and scale poorly with design size and workload length. Machine learning (ML)-based power estimation frameworks have shown promise in accelerating power estimation, but prior efforts only address average power or per-cycle power estimation. We propose \textbf{\textsc{\toolnamenospace}}, the first ML-based intra-cycle power estimation framework. \textsc{\toolnamenospace} operates purely on RTL simulation traces at inference time, eliminating the need for post-layout gate-level simulation and power analysis 
per workload. Across a diverse benchmark suite, \textsc{\toolnamenospace} achieves \textbf{5.88\%} median and \textbf{9\%} mean absolute percentage error compared to commercial post-layout gate-level power estimates while running \textbf{$\sim$80$\times$} faster. We further demonstrate that \textsc{\toolnamenospace}'s predictions can be reliably used for the downstream task of pre-silicon power side-channel leakage assessment.
\end{abstract} 

\vspace{-7pt}

\maketitle

\vspace{-7pt}
\section{Introduction}
\label{sec:intro}
Power consumption is a first-order design constraint in integrated circuits. Designers rely on fast and accurate power estimation throughout the design cycle to guide design decisions. Certain critical tasks further demand power estimates at \textit{intra-cycle} temporal granularity, i.e., at multiple instances within a single clock cycle. Power delivery network (PDN) design, for example, requires detecting peak current loads within a cycle to quantify voltage droop~\cite{Vakil2019IRATAIA} and perform timing closure~\cite{powernet}. Similarly, pre-silicon power side-channel analysis relies on input-dependent power signatures~\cite{dpa}. Fine-grained intra-cycle power analysis allows designers to reliably prevent information leakage by localizing where within a cycle the leak originates, allowing targeted mitigation. 

Existing power analysis 
tools span multiple levels of abstraction, with each level striking a different trade-off between accuracy, temporal granularity, and runtime. Lower-level flows of post-synthesis and post-layout deliver high accuracy and fine-grained temporal resolution because they capture the necessary physical details. 
However, this fidelity comes at a steep computational cost. Gate-level power evaluation requires logic synthesis followed by gate-level simulation (GLS), while layout-level analysis additionally requires place-and-route (P\&R), parasitic extraction, and SDF (Standard Delay Format) back-annotation of timing prior to post-layout GLS. Both flows scale poorly with design size and workload length, making them a bottleneck for large designs~\cite{gatspi,apollo,grannite}.

Machine learning-based power estimation has shown promise in alleviating these bottlenecks by improving the speed-accuracy trade-off~\cite{simmani,mlpower,grannite,apollo,deep,primal,agile}. Models are typically trained to map simulation traces or activity patterns at a higher level of abstraction (e.g., RTL) to power consumption labels collected from a lower level (e.g., gate-level netlist). However, prior work has addressed only average or per-cycle power estimation, leaving intra-cycle power estimation unexplored. We show that existing ML-based power estimation techniques do not directly apply to the intra-cycle setting.

\begin{figure}[!t]
    \centering
    \includegraphics[width=0.9\columnwidth]
    {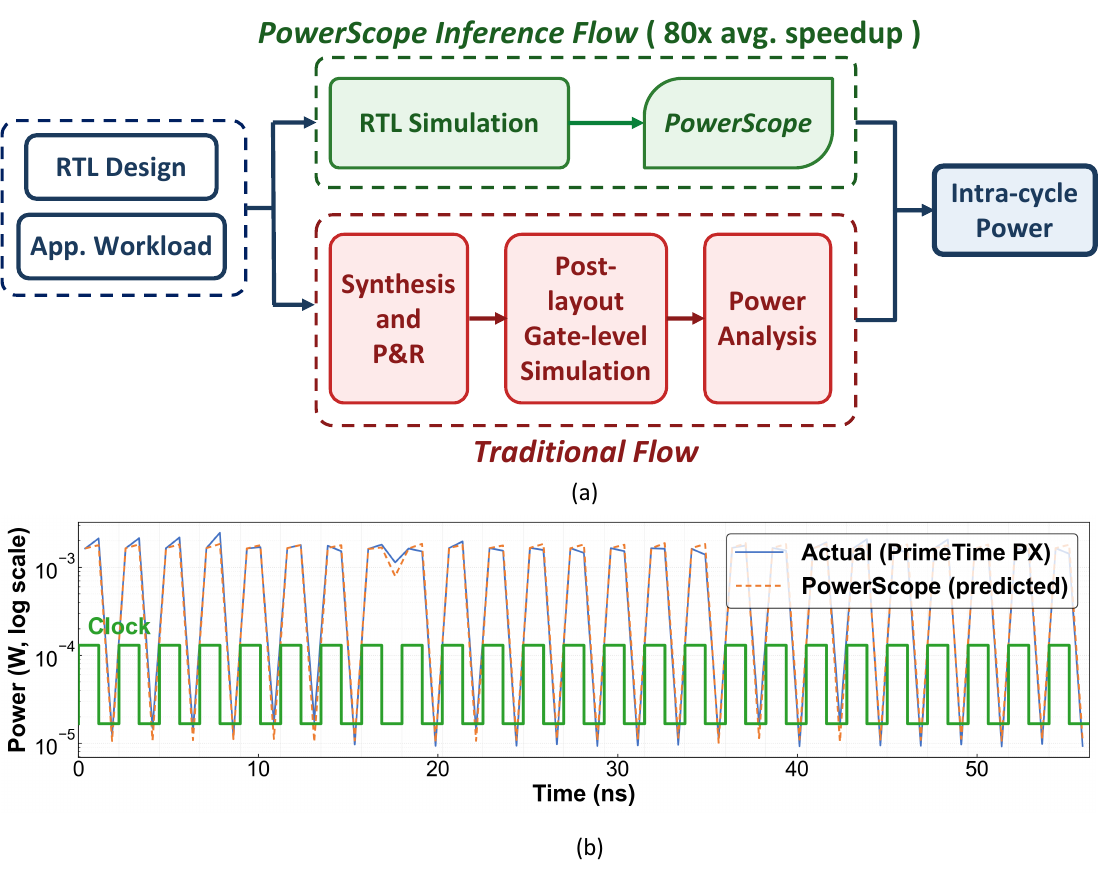}
    \caption{(a) \textsc{\toolnamenospace} Inference vs.\ the Traditional flow for intra-cycle power generation (b) Per-bin power comparison of \textsc{\toolnamenospace} prediction and Traditional flow, on the 64-bit Integer Radix-4 Divider~\cite{shakti} over 25 cycles. Speedup achieved is $\sim\!80\times$ end-to-end measured across designs. (excludes the one-time logic synthesis and layout cost)}
    \label{fig:comparsion_PowerScope}
    \Description{Comparison of \toolnamenospace inference and traditional flow, and per-bin power comparison on a 64-bit Integer Radix-4 Divider.}
\end{figure}
We propose \textsc{\toolnamenospace}, the first ML-based framework to estimate post-layout intra-cycle power directly from RTL simulation traces. Our key contributions are as follows:


\begin{itemize}[leftmargin=*, topsep=0pt]
    \item \textbf{Novel training methodology tailored to intra-cycle power.} We identify and address challenges such as imbalanced power distributions, that arise at finer temporal resolutions. We further propose an STA-guided feature selection technique that identifies the subset of RTL signals relevant to predicting power consumption at each interval within the clock cycle. 

    \item \textbf{RTL-only inference with post-layout power accuracy.} At inference time, \textsc{\toolnamenospace} operates entirely at the RTL level, eliminating post-layout GLS and power analysis for each workload. It achieves \textbf{9\%} mean absolute percentage error relative to the commercial post-layout power estimation flow while running \textbf{$\sim$80$\times$} faster.

    \item \textbf{Downstream application validation.} We demonstrate that \textsc{\toolnamenospace}'s intra-cycle power predictions enable a key downstream application of pre-silicon power side-channel analysis, with minimal accuracy loss.
\end{itemize}

\vspace{-7pt}
\section{Background}

\subsection{Intra-Cycle Power}
Let $T_{clk}$ denote the clock period of the design under evaluation. 
Power estimates can be obtained at various temporal resolutions. 
Average power estimation reports a single power value aggregated over an entire testbench or workload. Per-cycle power analysis reports one power value per clock cycle. Intra-cycle power analysis refers to the case where power is sampled at $B$ uniformly spaced temporal offsets within each clock cycle as illustrated in Figure~\ref{fig:power_dist} (a). Specifically, we divide $T_{\text{clk}}$ into $B$ {\em bins}, each spanning a sampling period $T_s = T_{\text{clk}} / B$ that sets the temporal resolution. Bin $b_i$ captures power at a fixed temporal offset $t_i = i \times T_s$ within the cycle, for $i \in \{0, 1, \ldots, B-1\}$.

\vspace{-7pt}

\section{Related Work}
\label{sec:relatedworks}
Power estimation is an important challenge in integrated circuit design, and has been extensively studied across various levels of design abstraction for over three decades. At the architectural and micro-architectural levels, tools such as WattWatcher~\cite{wattwatcher}, McPAT~\cite{mcpat} and McPAT-calib~\cite{mcpat-calib} provide fast estimates but operate at a coarse temporal granularity, reporting average power consumed across an entire application. At the RTL level, many early power estimation techniques used statistical and probabilistic approaches to accelerate power estimation, and were also limited to average RTL power~\cite{pre_ml_ssampling,najim,costa2004probability,costa1999,SequentialPowerEst,switchactivityRTL}. Other early works built linear regression models mapping hardware performance counter events to power~\cite{perfcounter} or mapping RTL switching activity to power consumed by RTL components~\cite{regressionrtlpower}, representing precursors to recent approaches to ML-based power estimation. Commercial signoff tools such as Synopsys PrimeTime PX~\cite{ptpx} provide accurate intra-cycle power estimates, but require synthesis, layout, post-layout GLS and power analysis to get intra-cycle power, making them subject to the scalability bottlenecks discussed in Section~\ref{sec:intro}. At the transistor level, FastSPICE~\cite{fastspice} simulators accelerate transistor-level simulation to achieve faster results, but remain computationally prohibitive for large designs~\cite{rewienski2011fastspice}.  

Recent years have seen growing interest in applying ML to improve the speed-accuracy trade-off in power estimation. One class of methods models power as a linear function of a small subset of RTL signals, which are referred to as \textit{power proxies}. SIMMANI~\cite{simmani} selects its power proxies based on mutual information between signal toggles and power consumption. APOLLO~\cite{apollo}, DEEP~\cite{deep} and ML-Power~\cite{mlpower} make use of MCP (Minimax Concave Penalty) regularization~\cite{mcp} during regression, which automatically selects power proxies while building  power models through regression. A second class of methods, including GRANNITE~\cite{grannite} and PRIMAL~\cite{primal}, uses neural network based models to learn mappings between switching activity and power consumption. These methods achieve better accuracy but incur a much higher cost to generate the large datasets required, and to perform training and inference. Recent works have shown that tree-based models are well suited for power estimation tasks, and even outperform deep learning approaches including Transformers and Graph-based networks in the context of RTL power prediction~\cite{masterRTL}. PRIMAL evaluates gradient tree boosting alongside neural networks, finding it competitive with neural networks in accuracy at significantly lower training and inference cost~\cite{primal}.


ML-based power estimation at intra-cycle temporal resolution remains unaddressed by all prior works. As discussed in Section~\ref{sec:intro}, intra-cycle estimates are essential for tasks such as PDN design, timing closure and pre-silicon power side-channel security analysis. Addressing this gap requires fundamentally new approaches to feature selection and modeling. \textsc{\toolnamenospace} fills this gap by proposing the first ML-based framework to achieve post-layout power accuracy at intra-cycle temporal resolution, while operating exclusively at the RTL level during inference.

\begin{figure}[t]
    \centering
    \includegraphics[width=0.85\linewidth]{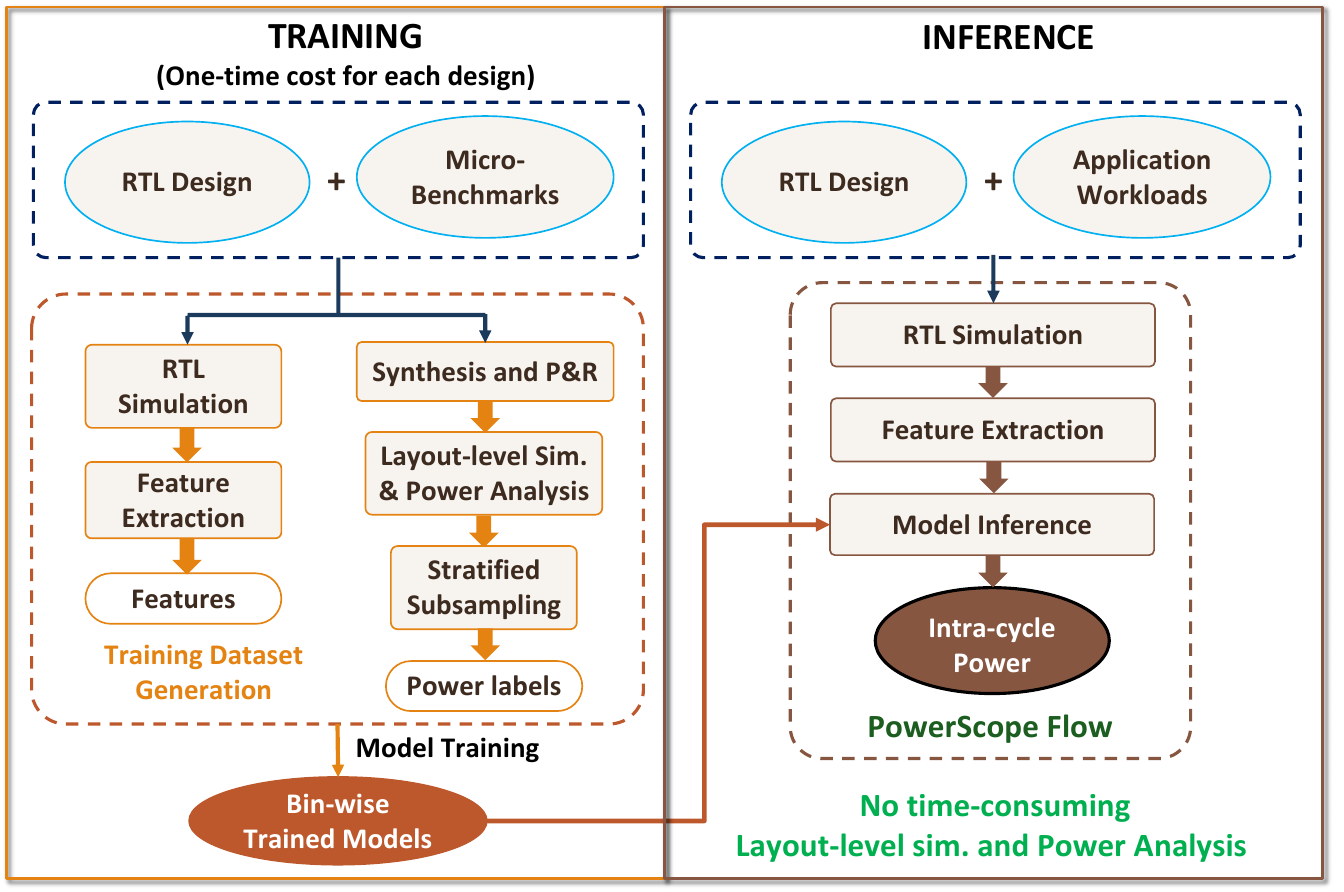}
    \caption{\textsc{\toolnamenospace} Training and Inference flow}
    \label{fig:PowerScope}
\end{figure}

\vspace{-15pt}
\section{Methodology}
\label{sec:method}

\textsc{\toolnamenospace} operates in two phases, as illustrated in Figure~\ref{fig:PowerScope}. During the one-time training phase, the  RTL design and a set of \emph{micro-benchmarks} drive two parallel pipelines. The first runs RTL simulation followed by feature extraction to produce the input features for training. The second runs logic synthesis, place-and-route (P\&R), and post-layout gate-level power analysis, followed by stratified subsampling to obtain ground-truth intra-cycle power labels. The training phase incurs a one-time cost of synthesis, P\&R, and post-layout GLS to generate the training dataset. 

During inference, \textsc{\toolnamenospace} requires only the RTL design and the application workload. The RTL is simulated to extract the required features, which are then fed to the trained models to produce the intra-cycle power trace, entirely bypassing the expensive step of post-layout gate-level simulation and power analysis. We describe the dataset generation, training and inference phases of \textsc{\toolnamenospace} in further detail in the following subsections.
\vspace{-5pt}
\subsection{Training Dataset Generation}
\subsubsection{Power Labels and Stratified Sub-sampling}
The ground-truth intra-cycle power labels are generated through a standard post-layout gate-level power estimation flow. First, the RTL design is synthesized to a gate-level netlist, followed by P\&R. Post-layout GLS is performed with back-annotated 
parasitics for a set of micro-benchmarks. The resulting traces in the form of Value Change Dump (VCD) files obtained from simulations are subsequently used as inputs to a power analysis tool. The power analysis tool also takes as input the desired sampling resolution $T_s$ and generates the ground-truth power labels for each of the $B$ bins within each clock cycle. 

\begin{figure}[t]
    \centering
    \includegraphics[width=0.85\linewidth]{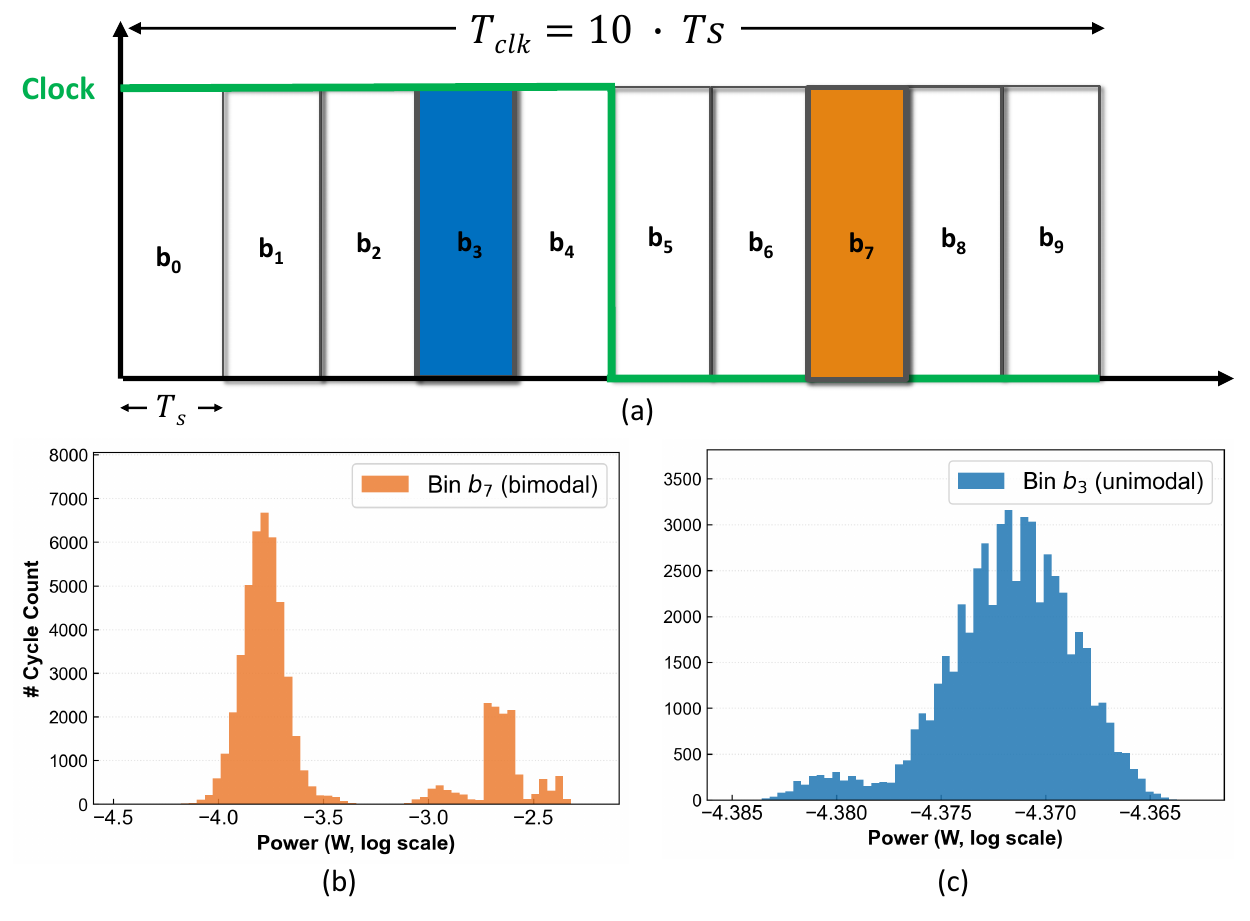}
   \caption{
 (a)~Illustration of temporal bins $b_0$ through $b_9$ within a single clock cycle. Histogram of per-bin power across 
60K simulated cycles for two temporal bins of an OpenTitan AES design~\cite{opentitan}. (b)~Bin $b_7$ is dominated by a leakage mode at low power, with a smaller activity mode at higher power, producing a 
bimodal distribution (c)~Bin $b_3$ exhibits a unimodal leakage-dominated distribution }
    \label{fig:power_dist}
\end{figure}

Most ML-based power estimation methods train on a  random subset of the simulated cycles, {\em i.e.}, perform a random splitting of the dataset into the training and validation or test sets. However, we observed that using random sampling to train power models for each bin leads to  accuracy degradation for bins with imbalanced power distributions. We attribute this to the highly skewed power distribution for certain bins as shown in Figure~\ref{fig:power_dist}. Bins that are very early in the clock cycle capture most of the signal transitions following the clock edge and tend to exhibit switching activity across most simulated cycles, producing a unimodal distribution centered around  higher power values (not depicted in Figure~\ref{fig:power_dist}). Some bins occurring later in the cycle exhibit no activity in most cycles, and also exhibit a unimodal distribution, but concentrated at leakage-level power as shown in Figure~\ref{fig:power_dist}~(c). In contrast, bins at certain temporal offsets within the clock cycle exhibit dynamic switching only under specific input patterns. As shown in Figure~\ref{fig:power_dist}~(b), this produces a bimodal distribution. It has a low-power mode corresponding to leakage-dominated cycles, and a separate higher-power mode corresponding to cycles with significant dynamic switching activity.

A naive random selection of cycles could undersample the leakage-dominated region and oversample the high-activity cycles, or vice versa, leading to poor coverage across the full power range during training. We address this using stratified subsampling to intelligently select a subset of cycles from the pool of simulated cycles. We apply stratified subsampling on a per-bin basis during training dataset construction. 

 For each bin requiring stratified subsampling, we sort the cycles by their power values and partition them into $K$ equal population strata. We then sample approximately $\lceil N/K \rceil$ cycles uniformly from each stratum where $N$ is the target training set size. If a stratum contains fewer than the requested number of cycles, the shortfall is redistributed proportionally to denser strata so that the target dataset size $N$ is met exactly. This quantile-based stratification ensures all  power regimes are well represented in the training set, regardless of how imbalanced the original distribution is.

Quantile-based stratification has the highest impact on accuracy for bins with bi-modal power distributions (Figure~\ref{fig:power_dist}~(b)), where the high activity cycles would be under-represented by naive random sampling. For such bins, the procedure substantially re-balances the training dataset, allowing for good coverage across the full power range. For other bins (Figure~\ref{fig:power_dist}~(c)), the per-quantile populations closely resemble those produced by random sampling, and the procedure has negligible effect.
\vspace{-3pt}
\subsubsection{Feature Selection}
\label{sec:proxy_selection}
Prior ML-based per-cycle power estimation works select a small subset of RTL signals as power proxies using L1-regularization methods such as the minimax concave penalty (MCP) to predict per-cycle power. MCP automatically down-selects a small set of power proxies from the full candidate set. However, extending this approach to intra-cycle estimation is impractical. MCP requires an iterative hyperparameter search to balance estimation accuracy against the number of selected proxies. At intra-cycle granularity, this search must be repeated independently for each of the $B$ bins in the design, significantly increasing training cost. 

We address this limitation with a static timing analysis (STA) guided feature selection method. Our method is based on the following observation:
Since gates have non-zero propagation delays, different gates switch at different temporal offsets within the clock cycle. STA can be used to derive a switching window for each gate. Gate $g$ contributes to the dynamic power of bin $b_i$ only if its switching window overlaps with the timing window represented by bin $b_i$. Thus, each bin $b_{i}$ has a distinct subset of gates, called the {\em potential switching subset} (PSS), that contribute to its dynamic power. The PSS for each bin is typically small relative to the total set of gates in the design, and is simply and efficiently constructed using the gate-level netlist and the STA timing report. 

We trace the transitive fan-in of each gate $g$ in the PSS of a bin back to registers, to obtain a subset of registers in the RTL that determine the switching activity in bin $b_{i}$. For each such selected register $R$ and each cycle $c$, we record two binary features that are derived from cycle-level RTL simulation. The first is the register toggle which indicates switching activity and the second is the state of $R$ in the previous cycle ($c-1$). Together, these features capture the input-dependent component of switching power. These per-bin feature sets constitute the input to the ML model. The full procedure for STA-guided feature selection is described in Algorithm~\ref{alg:sta_feature_selection}.

\begin{algorithm2e}[h!]
\caption{STA-Guided Feature Selection}
\label{alg:sta_feature_selection}
\KwIn{$G$ = Gate-level netlist, SDF (Standard Delay Format) file with delays, $B$, $T_s$}

\KwOut{$F_b$ for each $b \in \{0, \ldots, B-1\}$ 
\tcp*{Bin-specific feature set}}
\tcp{Step 1: Identify bin-specific Register set}
\For{each gate $g \in G$}{
    $t_{\min}(g), t_{\max}(g) \gets$ min/max arrival 
times at output of $g$ \tcc*{Derived from SDF}
    
    \For{$b \gets 0$ \KwTo $B-1$}{
        \If{$[t_{\min}(g),\, t_{\max}(g)]$ overlaps with bin $b$}{
            $r \gets \textsc{Fan\_in\_Registers}(g, G)$ 
            \tcc*{Trace back to registers driving the gate}
            $R_b \gets R_b \cup r \;$
            \tcp*{Per-bin Register set}
        }
    }
}

\tcp{Step 2: Build feature set from RTL simulation}
\For{each bin $b \in \{0,\ldots,B-1\}$}{
    \For{each register $R \in R_b$, each cycle $c$}{
        $\text{toggle}(R,c) \gets R(c) \oplus R(c-1)$
        $\text{state}(R,c) \gets R(c-1)$
        $F_b.\text{append}(\{\text{toggle}(R,c),\,\text{state}(R,c)\})$\;
    }
}
\Return $F_b$ for all $b \in \{0,\ldots,B-1\}$\;
\end{algorithm2e}

\subsection{Model Training}
\label{sec:training}

ML-based power estimation works predominantly employ 
linear models~\cite{simmani, apollo, mlpower, deep}, which 
express power as a weighted sum of selected RTL features. While computationally efficient and amenable to hardware implementation, linear models cannot capture non-linear effects inherent to gate-level power dissipation, such as glitch generation and propagation due to unbalanced path delays, and logic reconvergence. 

Deep neural network-based approaches~\cite{grannite, primal} achieve good accuracy but require large training datasets and incur significant training and inference costs~\cite{masterRTL}. Tree ensemble models provide a middle ground by being sample-efficient compared to neural networks while also capturing non-linear patterns effectively~\cite{masterRTL,panda,primal}. We hence adopt LightGBM~\cite{lightgbm} as our per-bin power model.

We train one independent LightGBM model per bin $b_i$. The input to each model is the bin-specific feature set $F_{b}$, derived from STA-guided selection, consisting of the register toggle indicator and register state for each selected register $R$. Since bins vary in their power distributions, we adapt the tree ensemble capacity on a per-bin basis. We perform a grid search over tree-ensemble hyper-parameters to pick the best configuration for each bin. Bins with high-activity use wide and shallow ensembles with a large number of leaf nodes, low depth and light regularization. This is attributed to many simultaneously toggling registers whose individual switching contributions combine to determine the bin's total power. Deeper and more strongly regularized ensembles are selected for bins that exhibit a wide dynamic range of power across cycles. For the bins which are dominated by leakage power, small heavily regularized ensembles are selected, since a low-capacity model can effectively predict the bin's power.

\vspace{-7pt}
\subsection{Inference}
During inference, \textsc{\toolnamenospace} operates entirely at the RTL level. Given an application workload, RTL simulation is first performed to produce cycle-by-cycle traces of all registers. The bin-specific STA-guided feature sets identified during training are then extracted from the resulting trace. These bin-specific features are fed to the corresponding LightGBM models for each bin to estimate intra-cycle power. \textsc{\toolnamenospace} thus accelerates intra-cycle power estimation by bypassing computationally expensive post-layout GLS and power analysis, which is only required during training.

\vspace{-7pt}
\section{Experiments \& Results}
\subsection{Experimental Setup}
\noindent\textbf{Tool Flow:}
All RTL and post-layout gate-level simulations are performed using Synopsys VCS~\cite{vcs}. Logic synthesis is performed using Synopsys Design Compiler~\cite{dc}, and place-and-route is performed using Cadence Innovus~\cite{innovus}, both targeting the GlobalFoundries 22nm technology node~\cite{gf22}. Ground-truth intra-cycle power labels are generated using Synopsys PrimeTime PX~\cite{ptpx}. All ML model training and inference experiments are conducted on an Intel Xeon server with 2\,TB of RAM and 512 cores.

\noindent\textbf{Designs Evaluated:}
We evaluate \textsc{\toolnamenospace} on designs of varying sizes and architectural complexities, spanning combinational, multi-cycle, and pipelined datapaths. These include arithmetic circuits from the EPFL benchmark suite~\cite{epfl}, a multi-cycle 64-bit integer radix-4 divider, a pipelined 64-bit floating-point fused multiply-accumulate (FMA) module used in an open-source RISC-V processor~\cite{shakti}, and the AES cipher core from OpenTitan~\cite{opentitan}. For each design, a representative set of micro-benchmarks are chosen that cover diverse switching patterns and power ranges to construct the training dataset. 

\noindent\textbf{Metrics:}
We report the following metrics for the selected designs, averaged across all $B$ bins: mean absolute percentage error (MAPE), median absolute percentage error (MedAPE), and Spearman correlation $\rho$. 
Spearman $\rho$ is computed by pooling the test set predictions from all $B$ bins into a single ranked list and measuring the correlation between the predicted and ground-truth power values. 
Absolute accuracy (MAPE and MedAPE) and rank correlation ($\rho$) capture different aspects of power estimation, and their relative importance is application dependent. Some downstream applications, such as pre-silicon power side-channel analysis, rely primarily on the trend of the power samples in the trace rather than the absolute accuracy~\cite{brier2004correlation}, whereas tasks such as PDN design and IR signoff require absolute power accuracy, because errors propagate into voltage droop and timing analysis~\cite{powernet,Vakil2019IRATAIA}.

\begin{table*}[!t]
\centering
\begin{tabular}{|c|c|c|c|c|c|c|c|c|c|}
\hline
\textbf{Design}  & \textbf{\#Cells} &
\textbf{Type} & \textbf{$T_{clk}$ (ns)} &
\textbf{$T_s$ (ps)} & \textbf{$B$} &
\textbf{Median APE} & \textbf{MAPE} &
\textbf{Spearman} $(\rho)$ & \textbf{Speedup}\\
 & &  & & & & \textbf{Average (\%)} & \textbf{Average (\%)} & \textbf{Correlation} & \\
\hline
  Fixed-point adder$^{\ast}$    &    635 & Arithmetic     & 7.50 &  750 & 10 &  2.40 &   2.79 & 0.276 &  15.2$\times$ \\
\hline
  Radix-4 divider$^{\dagger}$   & 7{,}746 & Arithmetic     & 2.25 &  750 &  3 &  9.53 &  14.81 & 0.818 &  83.3$\times$ \\
\hline
  FMA$^{\ddagger}$              & 8{,}910 & Arithmetic     & 10.0 & 2000 &  5 &  8.44 &  13.20 & 0.959 & 248.4$\times$ \\
\hline
  EPFL square$^{\ast}$          & 10{,}906 & Arithmetic     & 4.00 &  500 &  8 & 13.84 &  17.81 & 0.763 &  49.2$\times$ \\
\hline
  AES$^{\dagger}$               & 40{,}619 & Cryptographic & 5.00 &  500 & 10 &  2.64 & 6.27  & 0.945 & 207.6$\times$ \\
\hline
\end{tabular}
\centering
\captionsetup{justification=centering} 
\caption{Design characteristics and \textsc{\toolnamenospace} accuracy across 
 designs, evaluated at layout-level. \\ Design type: $\ast$~combinational, 
$\dagger$~multi-cycle, $\ddagger$~pipelined.}
\label{tab:results_overall}
\end{table*}
\vspace{-5pt}
\subsection{Results}
Table~\ref{tab:results_overall} summarizes \textsc{\toolnamenospace}'s accuracy across designs.
For all designs, metrics are measured on workloads completely unseen during model training.  Across designs, \textsc{\toolnamenospace} achieves an average MedAPE of \textbf{5.88\%} and an average MAPE of \textbf{9.05\%}.

The FMA design exhibits an average MAPE of $13.2\%$ alongside $\rho = 0.96$, indicating that the model reliably preserves the trend of power within the cycle. The combinational Fixed-point adder achieves the lowest absolute error across all designs (MAPE = $2.79\%$, MedAPE = $2.4\%$), averaged across all the bins, even though the $\rho$ is low ($0.276$). This can be attributed to the fact that in most cycles, the switching activity settles within the early bins (earlier as compared to other designs), while most of the bins occurring later in the cycle see no activity and are dominated by leakage power. Hence, there is limited relative power variation among the bins, leading to a low $\rho$ value that is not a meaningful metric in this scenario. 

\begin{table}[h]
\begin{tabular}{|l|c|c|c|}
\hline
\textbf{Design} & \textbf{\#Bins } & \textbf{\# Selected proxies }
                & \textbf{Total \#}  \\
&($B$) & \textbf{per-bin(min--max)} & \textbf{RTL Signals} \\
\hline
Adder      & 10 & 64--256     &  1{,}276  \\
Radix-4         &  3 & 75--138     &  9{,}389  \\
FMA             &  5 & 329--626    &  7{,}485 \\
EPFL Square          &  8 & 63--63      & 18{,}550  \\
AES             & 10 & 51--2{,}303 & 47{,}855 \\
\hline

\end{tabular}
\captionsetup{justification=centering} 
\caption{STA-guided feature selection selects 1--20\% of RTL signals across designs as power proxies}
\label{tab:sta_proxies}
\end{table}
\vspace{-3pt}
\subsubsection{Power Proxy Selection Speedup:}
Table~\ref{tab:sta_proxies} reports the number of power proxies per bin selected by the STA-guided procedure across all designs (described in Sec~\ref{sec:proxy_selection}). For the smallest design, no bin uses more than $20.1\%$ of the available RTL signals, while in the largest design, AES, the highest fraction of selected proxies just drops to $5.0\%$. Bin $b_0$ captures the rising clock edge, where in general the largest switching activity occurs, and hence $b_0$ consistently requires most proxies across all designs. Subsequent bins in the cycle require progressively smaller numbers of proxies. In the AES design, for instance, the count drops from $2{,}303$ in $b_0$ to $51$ in the last bin, $b_9$. 

\begin{figure}[t]
    \centering
    \includegraphics[width=0.85\linewidth]{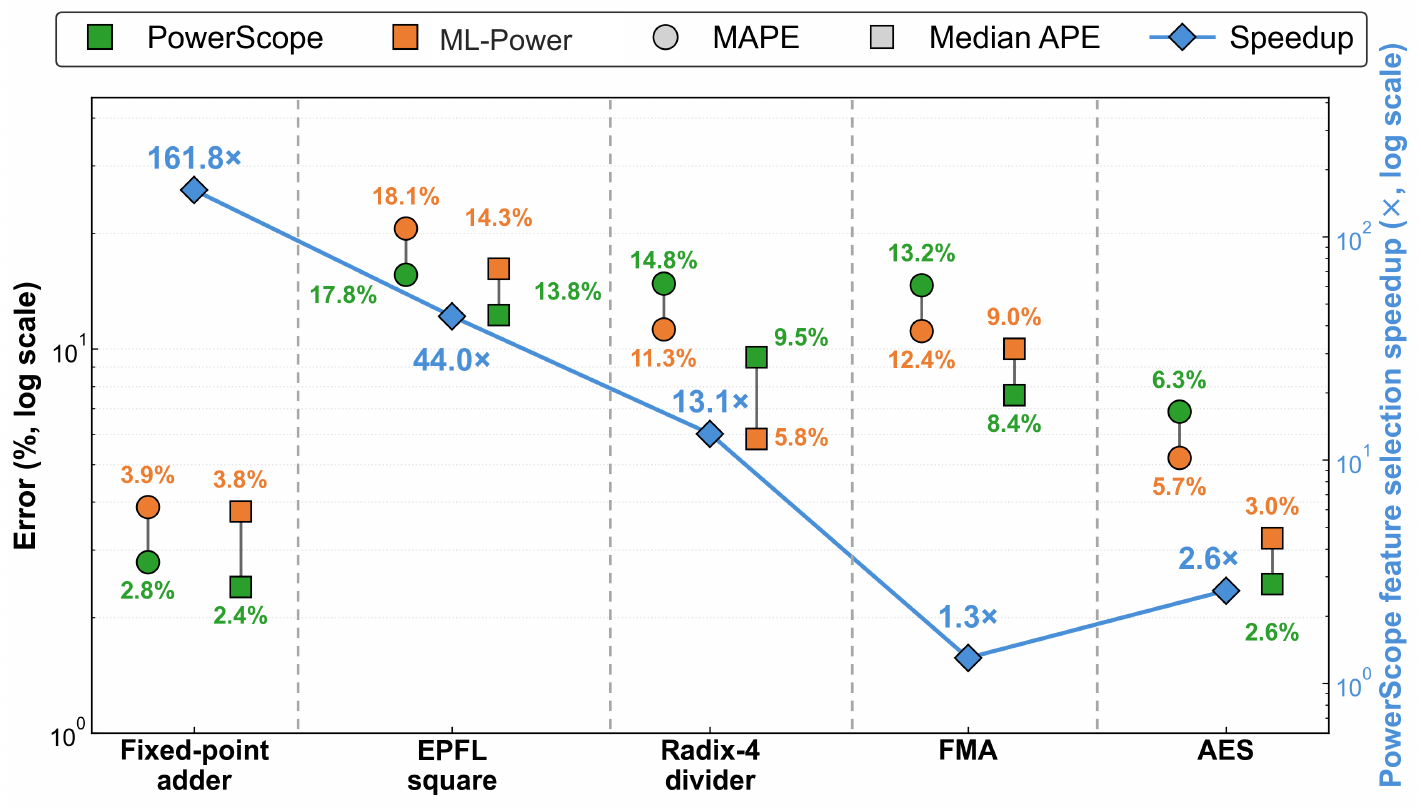}
    \captionsetup{justification=centering}     
    \caption{Accuracy and feature-selection speedup of \textsc{\toolnamenospace} vs ML-Power~\cite{mlpower}}
    \label{fig:ps_vs_mlpower}
\end{figure}

Prior ML-based per-cycle power estimators~\cite{apollo,deep,mlpower} use MCP-based proxy selection and require a separate iterative hyperparameter search for each of the $B$ bins, scaling feature-selection cost linearly with $B$. As the sampling period $T_s$ is made finer for higher temporal resolution, $B$ grows proportionally, making this per-bin MCP-based feature selection increasingly expensive. For a fair comparison of proxy selection time, for MCP-based methods, we perform hyperparameter search with 5 values of MCP penalty for each bin and select the model that performs the best with the least number of proxies.
\textsc{\toolnamenospace}'s STA-guided approach produces all the bin-specific proxies in a single pass over the gate-level netlist and SDF file independent of $B$. 
 Figure~\ref{fig:ps_vs_mlpower} shows the feature selection speedup (right y-axis) and prediction error (left y-axis)  for \textsc{\toolnamenospace} and \cite{mlpower}. \textsc{\toolnamenospace} achieves an average $12.5\times$ speedup over ML-Power across the five designs, while also being equally or more  accurate than ML-Power. 

\begin{figure}[t]
    \centering
    \includegraphics[width=0.85\linewidth]{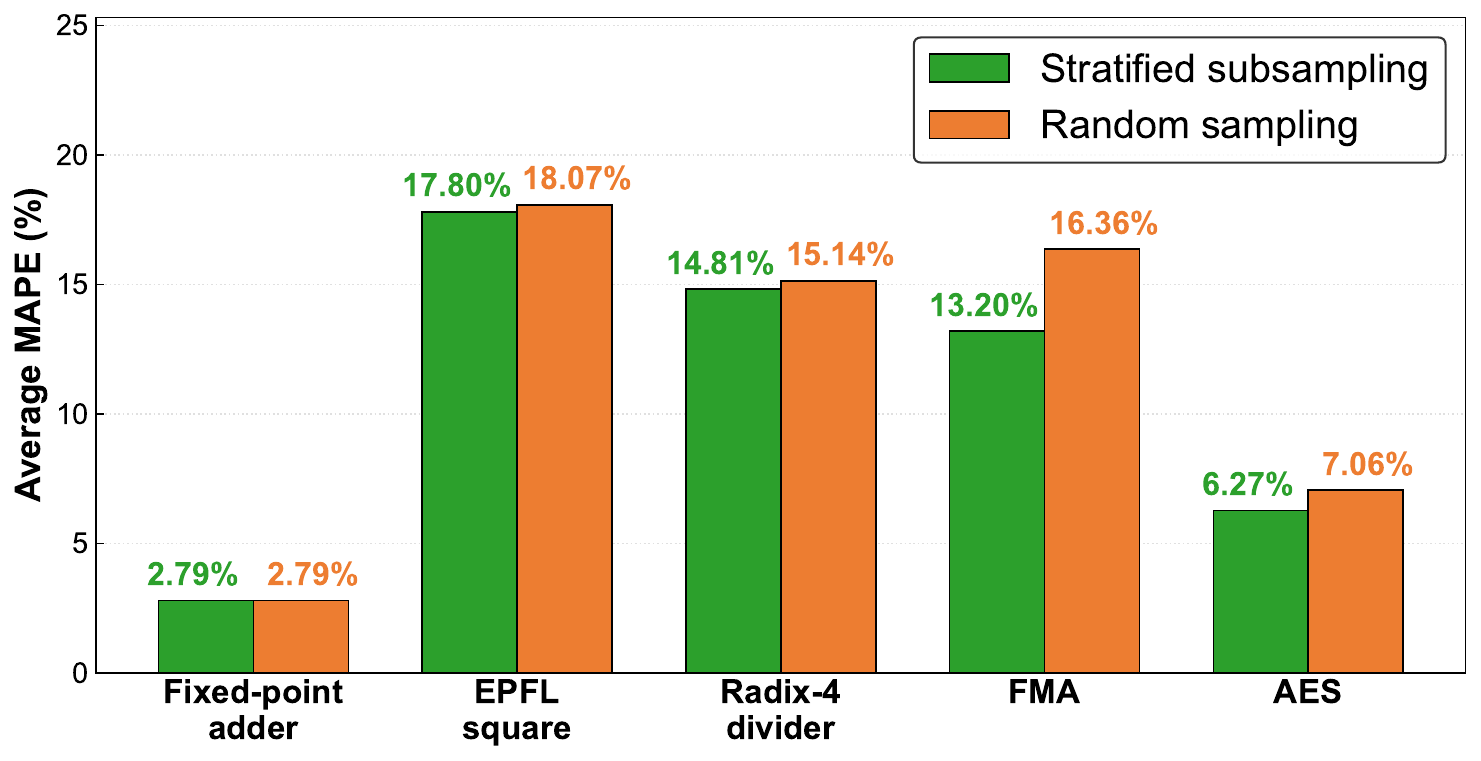}
    \captionsetup{justification=centering}     
    \caption{\textsc{\toolnamenospace}'s stratified subsampling (green) vs. random sampling (orange)}
    \label{fig:RandvsStrat}
\end{figure}

\subsubsection{Advantage of Stratified Subsampling:}
\textsc{\toolnamenospace} employs stratified subsampling, which ensures uniform coverage across the entire power range for each bin. To quantify its advantage, we compare it with random sampling.
Figure~\ref{fig:RandvsStrat} shows the comparison of the MAPE achieved by using both sampling strategies during training for all designs. All other components of the training flow remain unchanged.  Fixed-point adder, EPFL Square, and Radix-4 divider designs show little improvement for stratified subsampling over random sampling because these designs are dominated by bins with unimodal power distributions (similar to Figure~\ref{fig:power_dist}(c)). So, both sampling techniques achieve similar coverage of the power distribution. 
The FMA and AES designs, in contrast, contain more bins with a bimodal power distribution (similar to Figure~\ref{fig:power_dist}(b)). As a result, random sampling is not able to reliably achieve coverage of both power modes. For the FMA design, using stratified sampling, the bins having bimodal distributions show MAPE reductions of $3$-$9\%$ over random sampling.
Beyond accuracy, stratified subsampling also yields a higher Spearman $\rho$, indicating that it helps better preserve the relative ordering of power values across cycles. 

\begin{figure}[t]
    \centering
    \includegraphics[width=0.85\linewidth]{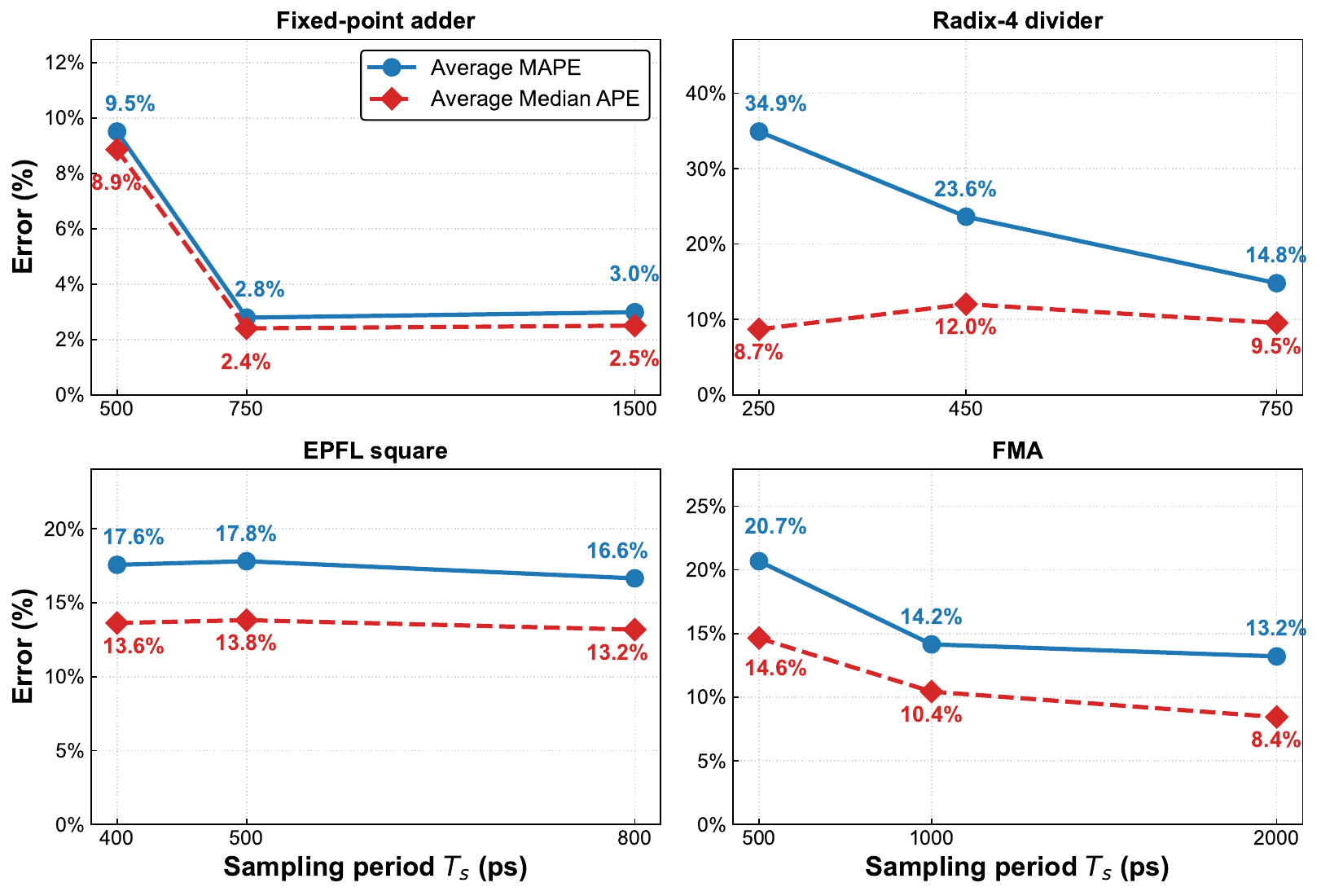}
    \caption{Effect of sampling period $T_s$ on prediction accuracy}
    \label{fig:ablation}
\end{figure}

\begin{figure}[t]
    \centering
    \includegraphics[width=0.85\linewidth]{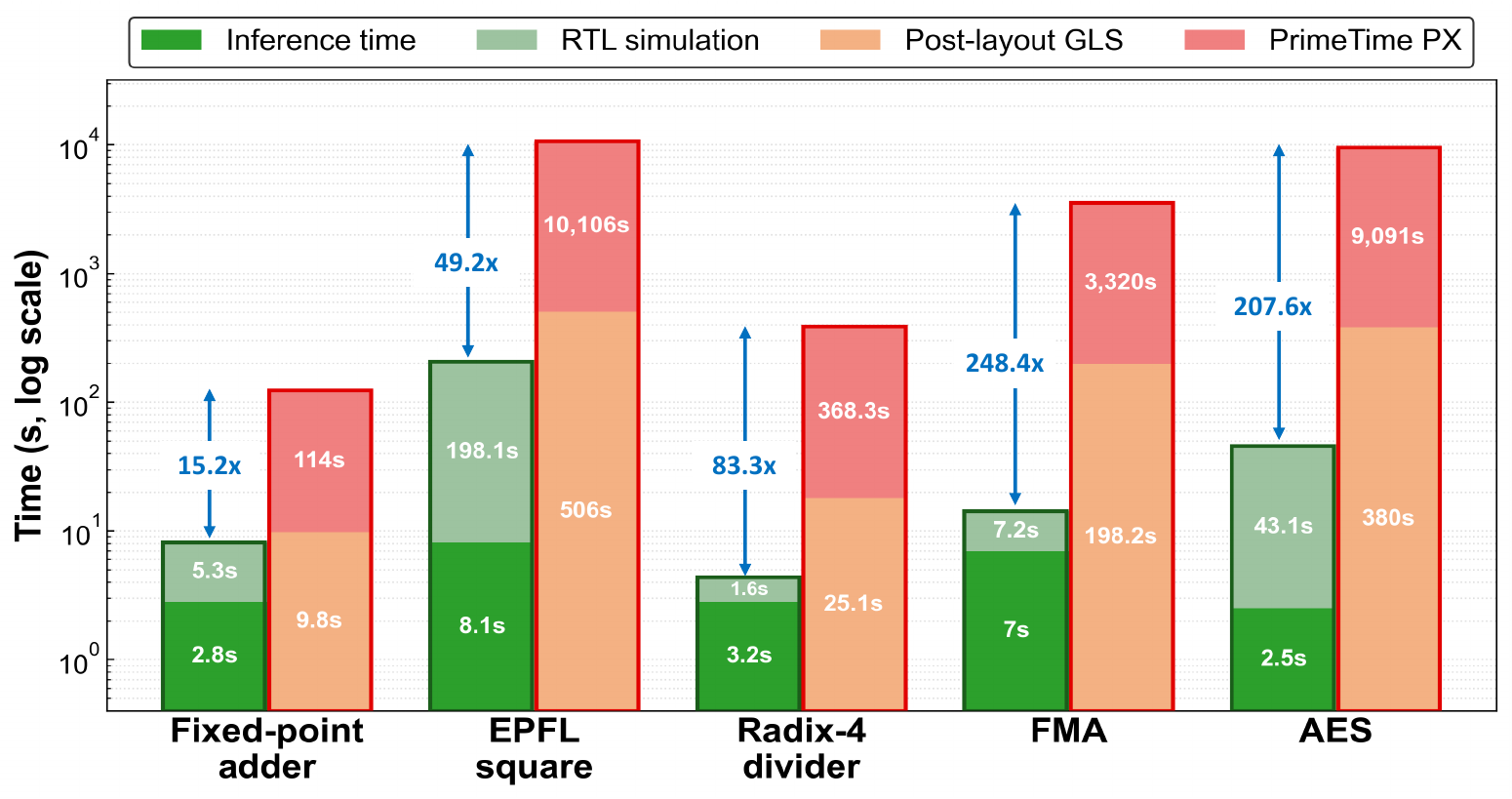}
    \caption{Runtime breakdown and speedup of \textsc{\toolnamenospace} inference (left bar per design) versus the traditional flow (right bar per design) for intra-cycle power across workload of 100k cycles. Speedup (annotated in blue) ranges from $15.2\times$ on the Fixed-point adder design to $248.4\times$ on FMA design. }
    \label{fig:PowerScope_speedup}
\end{figure}

\subsubsection{Varying Sampling periods:} Figure~\ref{fig:ablation} shows the effect of varying the sampling period~$T_s$ on prediction accuracy across designs.  For the Fixed-point Adder, Radix-4 Divider, and FMA designs, coarsening~$T_s$ reduces the prediction error. This is expected, as coarser bins pool power over wider temporal windows, compressing the per-bin dynamic range and yielding a smoother prediction target for the per-bin model.
However, as seen in the Fixed-point Adder design, this improvement can saturate beyond a point as it achieves the best accuracy at $T_s{=}750$\, ps with no further gain at $T_s{=}1500$\, ps.
The MAPE reduction with coarser~$T_s$ depends on how sharply concentrated each design's intra-cycle switching is within the clock period. Designs like the Fixed-point Adder and the Radix-4 Divider designs have a short toggle window after which switching activity subsides for the remainder of the cycle. At finer~$T_s$, the boundary between active switching and idle periods is divided into many narrow bins, each capturing only a fraction of the transition. In some cycles, these bins observe switching power while in others, only leakage power is captured, producing bimodal per-bin distributions that are inherently harder to predict at the RTL level. Coarsening~$T_s$ merges these narrow slices, averaging over the transition and yielding smoother per-bin power distributions. The EPFL square design, in contrast, is a combinational design in which switching activity ripples almost across the entire cycle, with no concentrated toggling. Most bins, therefore, have a similar kind of activity at various~$T_s$, and hence coarsening only marginally improves the model accuracy. 

\subsubsection{\textsc{\toolnamenospace} Inference speedup vs Commercial flow:}

Figure~\ref{fig:PowerScope_speedup} compares the end-to-end runtime of \textsc{\toolnamenospace} inference against the traditional post-layout power analysis flow across designs. The traditional power analysis flow comprises post-layout GLS (VCS) followed by power analysis (PrimeTime PX), while the \textsc{\toolnamenospace} flow includes RTL simulation followed by per-bin model inference run in parallel.
 In the traditional flow, both post-layout GLS and PrimeTime PX scale poorly with design size and workload length. \textsc{\toolnamenospace} eliminates these bottlenecks by replacing them with RTL simulation followed by per-bin model evaluation. The per-bin model's inference cost depends only on the number of selected proxies and the complexity of the tree ensemble (as discussed in Section~\ref{sec:training}).

\section{Downstream Application}
A motivating use case for fine-grained power analysis is pre-silicon power side-channel leakage detection~\cite{saidoyoki}. We use the Test Vector Leakage Assessment (TVLA) methodology described in ~\cite{goodwill}, which computes Welch's $t$-statistic for each bin. We evaluated the masked OpenTitan AES cipher core post-layout, by computing the TVLA $t$-statistic independently from PrimeTime~PX (ground truth) and \textsc{\toolnamenospace} (predicted). The \textsc{\toolnamenospace}-based $t$-statistic closely tracks the PrimeTime~PX ground truth, deviating by only $11.7\%$ on average across all bins. This confirms that \textsc{\toolnamenospace} enables rapid TVLA analysis directly from RTL, bypassing the time-consuming traditional power analysis flow. 

\begin{figure}[h]
    \centering
    \includegraphics[width=0.85\linewidth]{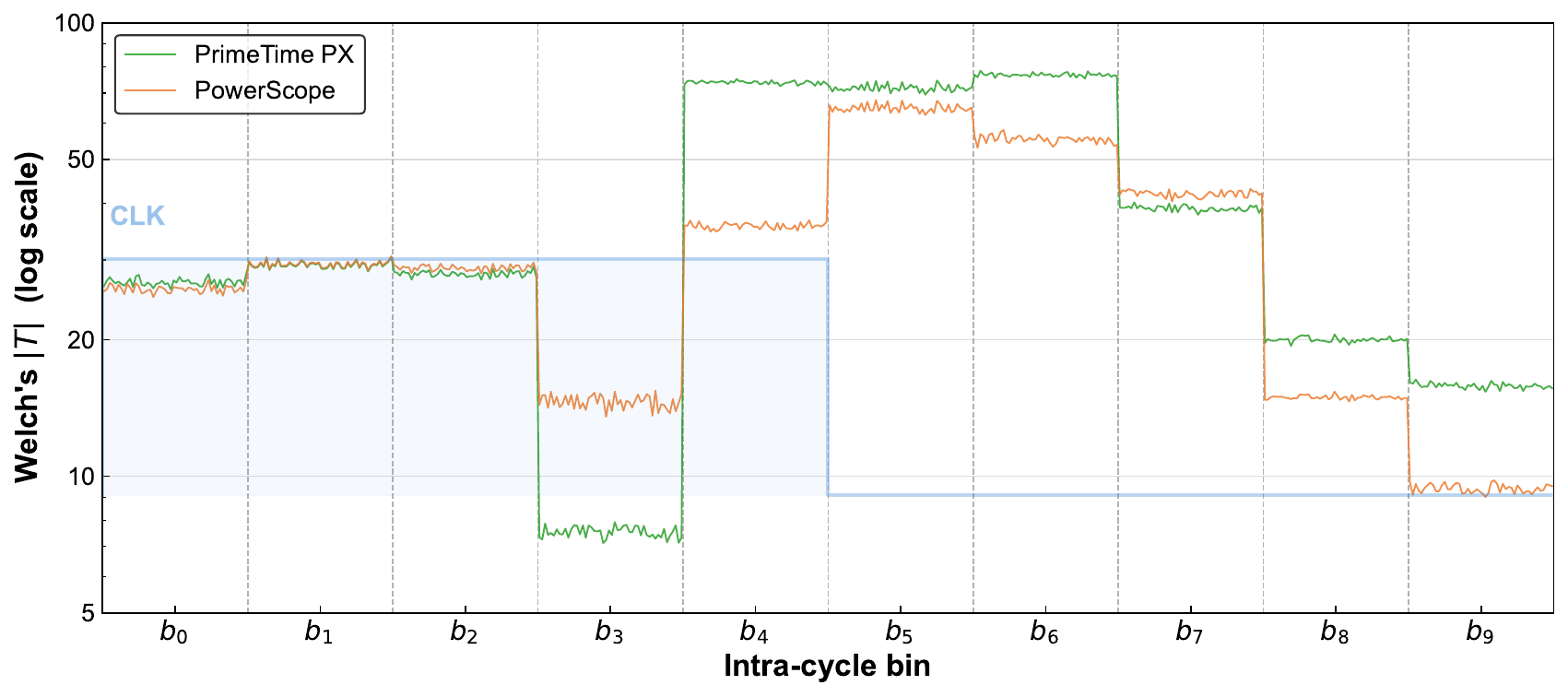}
    \caption{TVLA Welch's $|T|$ statistic per intra-cycle bin for the AES design ($T_s = 500$\,ps, $B = 10$), comparing PrimeTime~PX ground truth (green) against \textsc{\toolnamenospace} predictions (orange)}
    \label{fig:downstream}
\end{figure}
\vspace{-7pt}
\section{Conclusion}
We presented \textsc{\toolnamenospace}, the first ML-based framework to estimate intra-cycle power directly from RTL simulation traces. We proposed a novel training methodology that addresses the imbalanced power distributions that emerge at the intra-cycle temporal resolution through stratified subsampling. Our STA-guided feature selection identifies a compact, bin-specific set of power proxies, reducing the feature-selection cost by \textbf{${\sim}12.5\times$} over prior ML-based power proxy selection methods. \textsc{\toolnamenospace} achieves \textbf{$\sim$80$\times$} inference speedup over Synopsys PrimeTime~PX while estimating power to within $5.88\%$ median and $9\%$ mean absolute percentage error.  We further validated that \textsc{\toolnamenospace}'s intra-cycle power predictions transfer to the downstream application of pre-silicon power side-channel analysis using TVLA.

\section*{Acknowledgment}
This work was supported in part by Qualcomm. The authors would also like to thank Aradhana Mohan Parvathy for her valuable technical guidance and insightful discussions throughout this work.


\bibliographystyle{ACM-Reference-Format}
\bibliography{sample-base}

\end{document}